# OBSERVING NEAR EARTH ASTEROIDS
# WITH A SMALL TELESCOPE [1,2]

OVIDIU VADUVESCU

*York University, Department of Physics and Astronomy*
*128 Petrie Science Building, 4700 Keele St., Toronto, M3J 1P3, Ontario Canada*
*Email: ovidiuv@yorku.ca*

*Research Associate, Astronomical Institute of the Romanian Academy*
*Str. Cutitul de Argint 5, Bucharest RO-040557 Romania*
*Email: ovidiuv@yahoo.com*

*Abstract*. Even from a light polluted city it is possible to observe Near Earth Asteroids (NEAs) at opposition using a small telescope equipped with a CCD camera. In this paper, we will overview first the major NEA programs, continuing with planning the observations and the data reduction. Second, we will present a NEA follow-up program carried out on the 60-cm telescope at York University Observatory in Toronto, Canada. Part of this program, five NEAs have been observed during ten nights. Their astrometric and photometric data were reduced and sent to the Minor Planet Centre, following which an observatory code was assigned and four batches have been included in the NEODyS database and MPC Circulars. The results are applicable to any other small facility.

## 1. INTRODUCTION

Follow-up observations of Near Earth Asteroids are welcomed by the international astronomical community, in order to secure and improve their orbits, predict future close encounters with the Earth, and prevent any possible collision threat well ahead in advance.

Near Earth Asteroids (NEAs) are defined as the asteroids with a perihelion distance q<1.3 AU (NASA/JPL 2004a). Potentially Hazardous Asteroids (PHAs) are the NEAs having a Minimum Orbital Intersection Distance (MOID) less than 0.05 AU and absolute magnitudes H<22, which corresponds to objects larger than about 1 km. This limit in size represents the asteroids large enough to potentially cause a global climate disaster and threaten the continuation of human civilization (e.g., Chapman and Morrison 1994).

Conform to ASTORB database (Bowell 2004), there are 270,233 catalogued asteroids known today (Dec 16, 2004). Accordingly to the NEO website (NASA/JPL 2004a), there are 3108 NEAs and 660 PHAs known today (Dec 14, 2004). Although the annual increase has declined in the last two years, suggesting that there are fewer unknown objects detectable in the range of the present facilities, these two numbers continue to grow, with an annual number of new discoveries between 200-500 new NEAs and 50-90 new PHAs in the last 7 years (EARN 2004).

---

[1] Using data taken on the 60-cm telescope at York University Observatory, Toronto, Canada.
[2] Published in Romanian Astronomical Journal, Vol. 14, No. 2, 2004 and Vol. 15, No. 1, 2005.

Although most of the NEAs and PHAs are faint, having absolute magnitudes H>18.0 (corresponding to sizes smaller than ~1 km), about 24% of the NEAs and 22% of the PHAs have H<18.0, with 57 NEAs and 7 PHAs reaching H<15 (corresponding to sizes larger than ~5 km).

## 2. DISCOVERY, RECOVERY AND FOLLOW-UP PROGRAMS

The original aim of the international Spaceguard Foundation was to promote and coordinate activities for the discovery, follow-up and orbital calculation of the NEOs at an international level (Morrison 1992, Spaceguard 2004). To achieve these aims, NASA implemented a more specific goal to discover 90% of the NEAs greater than 1 km diameter in about one decade, by 2008 (e.g., Pilcher 1998). Part of this work-frame, there are about seven dedicated NEAs programs currently in progress: LINEAR (MIT 2004, currently holding more than half of the NEAs discoveries), NEAT (NASA/JPL 2004b), Spacewatch (LPL 2004), LONEOS (Lowell Observatory 1996) – all in the US, Catalina Sky Survey in the US and Australia (Beshore et al 2004), CINEOS in Italy (Boattini et al 2004a), and BAO in Japan (JSA 2004).

Two important steps after a NEA discovery are the recovery and follow up of the newly discovered object. Despite the increasing number of the NEAs, very few groups worldwide run dedicated follow up astrometry programs. A paper including a review about the astrometry of NEOs using small telescopes can be found in Steel and Marsden (1996). A more recent workshop on collaboration among NEO observers was carried out in Japan (JSA 2001).

Some notable follow-up efforts in the past were carried out at Dominion Astrophysical Observatory in Canada (Tatum et al 1994), McDonald Observatory in the US (Whipple 1995), the AANEAS program in Australia (Steel et al 1998), OCA-DLR in France and Germany (ODAS 1999), Beijing Observatory in China (Zhu 2000), also by D. Tholen on the Mauna Kea.

Today, among the most active NEAs follow-up groups are the Ondrejov and Klet Observatories in the Czech Republic (Pravec et al 2004, Tichá et al 2004, Tichá et al 2002, Pravec et al 1994), Modra Observatory in Slovakia (Astronomical Observatory Modra 2005), Pulkovo Observatory in Russia (Pulkovo NEO Page 2004), the US Naval Observatory using the FASTT scanning transit telescope (e.g., Stone 1997), and others. A recent pilot test to search and follow-up NEAs beyond 22nd magnitude was performed by Boattini et al (2004b) using two larger facilities, namely the 2.2m MPG/ESO telescope for the search and the 3.6m ESO NTT for the follow-up.

On the amateur and educational sides, is has been proven that small telescopes can be used successfully for follow-up of asteroids and NEAs by amateurs or at small colleges (e.g. Penhallow 1986, Oksa 2001). A very active amateur group involved in the follow-up observations, contributing with thousand of observations and many asteroid discoveries in the last decade is Sormano Observatory in Italy, using a 50-cm telescope (OAS 2004). A promising private automate search and follow-up project, MOTESS – the Moving Object and Transient Event Search System, employing three 35-cm telescopes with CCDs is under development by an amateur in the US (Tucker 2004a, Tucker 2004b).

Giving the growing number of known NEAs, more work can be done in the follow-up direction at the bright end by modest facilities, such as colleges and university observatories, also astronomy clubs and small amateur facilities, places where competition for observing time is not a problem, as it is on larger telescopes which can be spared for deeper studies.

# 3. PLANNING THE OBSERVATIONS

## 3.1. THE JPL NEO LIST

The NASA Near-Earth Object Program (NASA/JPL 2004a) maintains a list of future close approaches of Near Earth Objects (NEOs), including the approaching dates, miss distances, estimated diameters and relative velocities, on a period of about three months ahead. An additional database including all known NEAs and PHAs can be queried based on the absolute magnitude and the nominal distance. These references can be used in order to plan NEA observations close to the dates of their oppositions, when they reach their maximum brightness.

The exposure time to image a NEA at opposition is limited to the length of time that it takes to cross one pixel. For example, because of the excessive light pollution over the Greater Toronto Area, we could reach only a limit magnitude V~15 exposing for about half a minute.

Using the list of future close approaches (NASA/JPL 2004a), one can select about every three months the targets at opposition brighter than the observable limiting magnitude. Supposing an average miss distance of about 0.1AU, also our apparent limit magnitude V~15, one can calculate the diameter of a NEA in km, using the following classic formula (e.g. Harris 2001):

$$D = 1329 \times \frac{10^{-H/5}}{\sqrt{p_V}} \qquad (1)$$

Here $D$ represents the NEA diameter (considered spherical), $H$ is the absolute magnitude (defined as the magnitude of the object at 1 AU and a phase angle zero), and the $p_V$ is the albedo, which can be assumed to be 0.3 as an average value.

The relation between the apparent magnitude $m_V$ and the absolute magnitude $H$ is:

$$m_V = H + 5\log(r \times d) + P(\alpha) \qquad (2)$$

Here $d$ and $r$ represent the distances of the asteroid from the observer and the sun, respectively, and $P(\alpha)$ the "phase relation", which can be approximated by 0.3 at a phase angle zero. Plugging in equation (2) $m_V = 15$, $r = 1$AU, $d = 0.1$AU, $P(\alpha) = 0.3$, it follows $H = 19.7$, then from (1), using $p_V = 0.3$, we get $D = 0.3$ km. That is, any NEA larger than 300m satisfies these average conditions, becoming a potential target with an apparent magnitude $m_V<15$.

It is difficult to establish the frequency of the close approaches given this size limit for the whole NEA population, given the various possible geometry circumstances, albedos and sizes. Based on our two year experience, NEAs reaching V<15 with estimated diameters of about ~1km can be observed with a frequency of about one object every two months.

Once a list of possible targets is selected based on their size, one should run their daily ephemerides for one month around the close approach date, in order to check the observable circumstances. Various software or online servers can be used for the task, and we preferred HORIZONS System (JPL 2004).

## 3.2. EXPOSURE TIME AND PROPER MOTION

As we have shown above, close approaches of the faint NEAs are the only circumstances to observe such small bodies with a small telescope, especially from a bright place. Given these conditions, the exposure time at opposition has to be decided by "trading" the asteroid brightness versus its proper motion. While NEAs are the brightest at opposition, their proper motion is largest, and thus, the exposure time has to be reduced accordingly. That is, both the proper motion of the asteroid and its apparent magnitude must be taken into account for planning the observation of a NEA at opposition.

For example, assuming a pixel scale of 0.5 arcsec/pixel, an asteroid having an apparent motion of 0.1 arcsec/s could be exposed only for maximum 5s, in order to appear as a point and not as a trail. In practice the exposure time can go a little longer, because the asteroid image is blurred by the seeing (about 3 arcsec in Toronto). A few days before opposition (at elongations greater than 90 deg), the proper motion is smaller, thus the exposure time could be increased in order to reach its faint brightness.

## 3.3. EPHEMERIDES AND FINDING CHARTS

Precise finding charts and accurate ephemerides (to about 1 arcsec) are necessary to observe faint NEAs, especially for a small field of view and/or a dense stellar region.

There are two possible options to provide accurate ephemerides, either directly from a charting software or by importing them from an independent ephemerides software or an online server. Given the high accuracy needs, we preferred the HORIZONS server (JPL 2004) from which we imported the asteroid ephemerides into our own Celestial Maps v.9 charting software (Vaduvescu et al, 2002). In Fig. 1 we give an example of such a chart which we used in one of our observing nights. The bright object crossing the 0.5 deg chart in only five hours is PHA 65803, observed at York on 20/21 Nov, 2003. Other choices would be for example the IMCCE online server (IMCCE 2004) or the OrbFit client software (Milani 2004b). All three take into account planetary perturbations based on modern planetary theories (DE405/406 of JPL or VSOP87 of BdL), generating accurate NEAs ephemerides to about 0.1 arcsec.

An alternative to external ephemerides would be a charting software able to provide accurate ephemerides at any epoch, on its own. We recently added such a feature to Celestial Maps v.10 (Vaduvescu et al, 2005), based on the OrbFit numeric integrator code (Milani 2004b). For epochs close to osculation (~1 year), any other software providing accurate small field charts (field of view ~ 5 arcmin) working with very large star catalogs could be used instead. Three such packages are: XEphem (Downey 2004 – querying GSC 2.2 from an XEphem, STScI or ESO server, also USNO A and SA offline), Guide (Project Pluto 2004 – querying GSC 2.2 from STScI) and Astrometrica (Raab 2004 – querying USNO-B1 via VizieR). Should the observer choose to use other ephemerides software, we strongly recommend them to be checked against well tested ephemerides providers (e.g., HORIZONS).

Alternatively to client software, an online star chart service such as Aladin (CDS 2004a) could be used with various catalogs, although this does not plot yet the asteroid apparent path. We have used for the asteroid ephemerides the HORIZONS ICRF/J2000.0 astrometric positions calculated with the observer location set at topocentre (York University Observatory). We imported the ephemerides into Celestial Maps 9, plotting stars at epoch J2000, as the observing epoch is close to J2000.

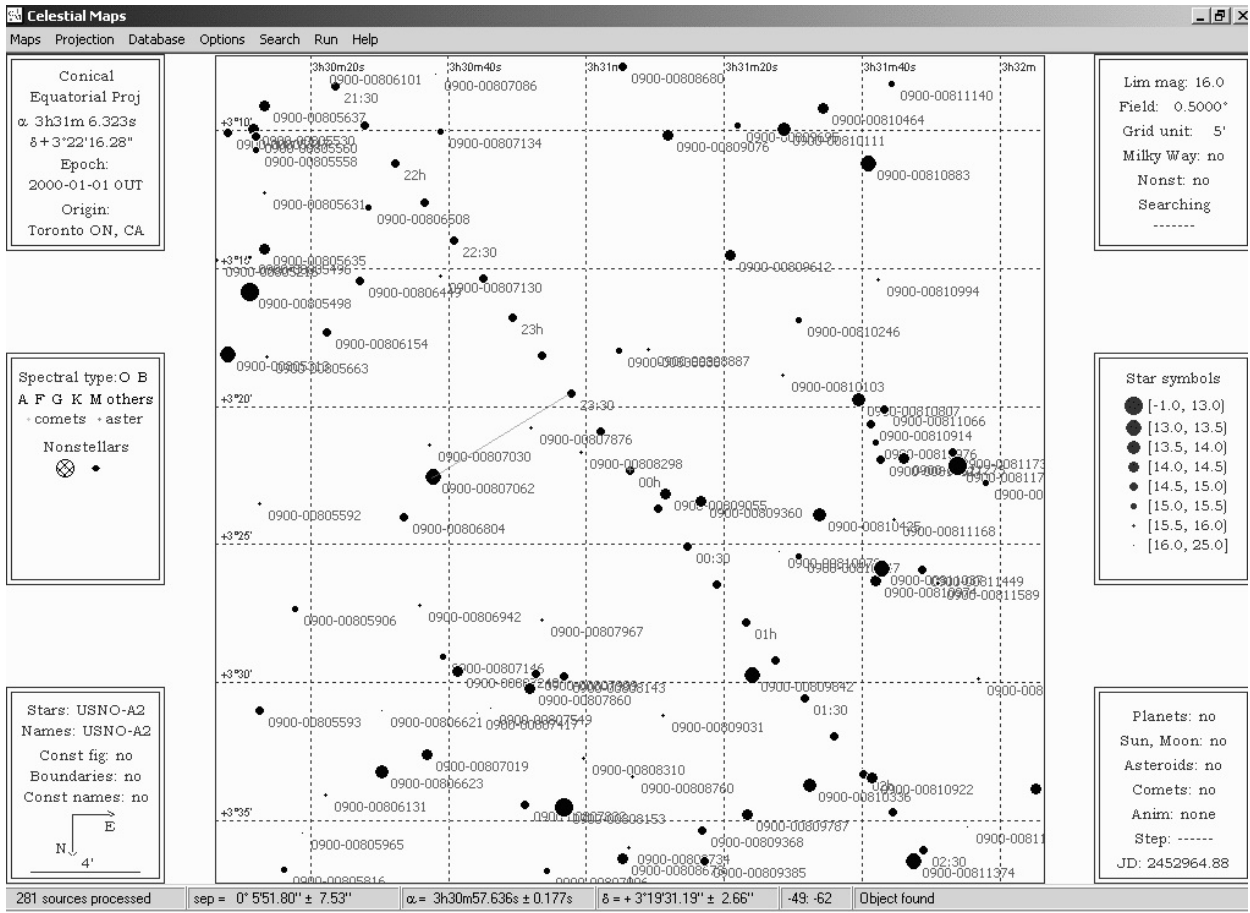

Fig.1 – Sample finding chart run with Celestial Maps v.10 using the USNO-A2 catalog and the HORIZONS ephemerides. The observing fields and catalog stars have to be chosen based on the NEA apparent path and the small FOV. The PHA 65803 was observed on Nov 20/21, 2003 at York, and can be seen crossing the field from SW to NE.

It is recommended to choose the observing fields and the reference stars in advance, based on the nightly apparent path of the asteroid through the stars plotted on a finding chart (e.g. every 15 minutes). Only catalog stars close enough to the asteroid can be imaged with a small field of view (FOV), and each field is desirable to have as many stars as possible. Therefore, all available catalogs should be checked before, in order to decide the best reference stars to be imaged in each field.

### 3.4. STAR CATALOGS AND THE FIELD OF VIEW

The main problem using a small CCD array (such us our 512x512 pixels) mounted on a long focal length telescope (such as our F/D=13.6) is the small field of view of the resulted image. This can become a major problem to observe asteroids from a light polluted place, where relatively few stars can be imaged given the short exposure time required by the fast moving NEAs at opposition.

Table 1 presents the main catalogs used nowadays for astrometry purposes, together with their main characteristics. Column one includes the catalog name, column two the total number of stars in the catalog (with the exception of USNO-B1, which lists the number of objects), column three the completeness (approximate stellar limit magnitude), column four the mean error in positions, while the last two columns give the star density per square degree, respectively the density in a small FOV 5'x5' similar to our facility at York Observatory.

*Table 1*

Star catalogs available nowadays for astrometry and their main characteristics

| Catalog | Nr of stars | Completeness | Mean error astrometry (") | Star density (per $deg^2$) | Star density (per 5'x5') |
|---|---|---|---|---|---|
| PPM | 468,861 | V~10.5 | ~ 0.3 | 11.4 | 0.08 |
| Tycho-2 | 2,557,501 | V~11.5 | ~ 0.06 | 62.0 | 0.43 |
| GSC 1.1 | ~15,000,000 | B~15.5 | ~ 1.5 | 456.7 | 3.17 |
| GSC 2.2 | ~ 178,000,000 | B~19.5 | ~ 0.2-0.4 | 11,050 | 76.74 |
| 2MASS | 470,992,970 | K~15.5 | ~ 0.12 | 11,417 | 79.28 |
| USNO-A2 | 526,280,881 | V~20.0 | ~ 0.25 | 12,757 | 88.59 |
| USNO-B1 | ~1 million objects | V~21.0 | ~ 0.20 | 25,354 | 176.07 |

The PPM catalog (Roeser and Bastian 1989, Bastian and Roeser 1992) provides relatively good astrometry (0.3 arcsec) and proper motions for a relatively low number of stars observed more than two decades ago, giving an average density of only 0.08 stars per CCD field, which is insufficient for a small CCD FOV.

Tycho-2 (Høg et al 2000) is a modern reference star catalog observed by Hipparcos Space Astrometry Mission at an average epoch 1991.5. The positions are given in the modern ICRS reference system as defined by Hipparcos catalog at epoch J2000. Having the best astrometric accuracy (0.06 arcsec) and including proper motion for 96% of its stars, Tycho-2 represents the most recommendable reference to use. Nevertheless, its stellar density is limited to about 0.4 stars per CCD field, thus being impractical to use alone in a small CCD field such as ours.

The GSC version 2 (STScI 2003) is an all-sky catalog based on 1 arcsec resolution scans of the photographic Sky Survey plates observed at two epochs and three bandpasses with the Palomar and UK Schmidt telescopes in both hemispheres. It was built to support operations of very large telescopes, such as the GEMINI and VLT. Its predecessor, "The Guide Star Catalog" (GSC 1.0), was released in 1990 with the main purpose to guide the Hubble Space Telescope in its space operations. The stellar positions in GSC 2.2 are observed at various epochs, and do not include proper motions. With the total number of objects including about 39% stars and 61% non-stars (STScI CASB, 2004), a careful star selection has to be made based on the object class flag.

2MASS Point Sources Catalog (Cutri et al, 2003) is a modern all-sky catalog of stars observed between 1996 and 2000 in the near infrared, on twin facilities in both hemispheres. It provides J2000 positions, good astrometric accuracy (~0.12 arcsec) with no proper motion, and a high stellar density (79 stars per CCD field), therefore being suitable to be use for small CCD field astrometry.

The USNO-A2 catalogue (Monet et. al, 1998) is based on a re-reduction of the Precision Measuring Machine (PMM) scans from two major archives at various observing epochs. It uses the modern ICRF reference as standard. Having a high density (89 stars per CCD field) but no proper motions, it provides a good astrometric reference for small CCD fields.

The USNO-B1 catalogue (Monet et. al, 2003) includes more than one billion objects whose positions were taken from scans of Schmidt plates from various sky surveys during the last 50 years. The catalog is expected to be complete to V~21, but not all sources in USNO-B1 are stars. The star/galaxy separation in USNO-B1 represents a measure of the similarity of the point-spread function to a stellar profile, as measured in each band. It is given by some flags, giving a number between 0-11, with a code closer to 0 meaning a galaxy-like object, while one closer to 11 a stellar-like object.

Celestial Maps software (Vaduvescu et al, 2005) is distributed on a CD-ROM which includes the PPM database (packing the four original PPM catalogs), GSC 1.1 database, and Tycho-2 database (including the two original Tycho-2 catalogs). The databases preserve the original catalog accuracy, indexing the stars in multiple files, based on the GSC architecture, coded for faster access. Starting with version 10, new online capabilities were added for automate FTP access to the large catalogs 2MASS, GSC 2.2, USNO-A2, and USNO-B1, queried via VizieR server (CDS 2004b).

## 4. THE FACILITY AT YORK UNIVERSITY OBSERVATORY

We have employed the 60-cm F/D=13.6 Cassegrain telescope, the largest of the two equatorial reflectors at York University Observatory (York Observatory, 2004). The two dome facility is located at the Department of Physics and Astronomy of York University (Keele campus) in Toronto, Ontario Canada. Its coordinates are: longitude 79d30'00" West, latitude 43d46'30" North, altitude 196m, MPC observatory code H79.

Initially, the telescope was equipped with a HPC1 512x512 pixels camera providing a field of view (FOV) of 4.86'x4.86' (0.56 arcsec/pixel). In August 2004, this CCD was retired and replaced with an SBIG ST-9 512x512 pixels camera, providing a FOV of 3.5'x3.5' (0.41 arcsec/pixel). Alternatively, this small FOV could be increased to 5.5'x5.5' or about 10'x10', using two focal reducer lenses. Nevertheless, despite of the advantage of having a larger field, we have some doubts about the possible distortions due to the field curvature introduced by an additional lens (e.g., Fierro and Calederon 2002).

The time stamp has been provided by the WinSNTP software (Coetanian Systems 2001) which uses the Simple Network Time Protocol (SNTP) to adjust the PC clock via a Winsock TCP/IP Internet query of a NTP time server (time.chu.nrc.ca), once every minute. Accordingly to its manual, taking into account the network delay between client PC and the time server, WinSNTP provides an accuracy of about 0.1-0.2s, sufficient for our purpose. We check this accuracy each night, using the USNO web master clock, also the time signal phone service provided by the National Research Council in Ottawa. The time stamp is recorded in the FITS image headers with an accuracy of 0.1s by the HPC1 camera and 0.5s by the ST-9 camera.

A Pentium II Intel MMX PC at 400 MHz with 200 MB RAM running Windows 95 and recently Windows 98 has supported the data acquisition. We have used the DEVON software to run the HPC1 camera, and more recently CCDOPS to command the ST-9 camera.

For the photometry purposes, a regular Johnson BVRI system filter was used manually with the HPC1 camera, and an automate filter wheel with the ST-9 camera.

## 5. THE OBSERVATIONS

Between Nov 2002 and Sep 2004, we observed 5 asteroids over 8 nights, namely PHA 35396, NEA 40267 (2 nights), PHA 65803 Didymos, NEA 2002 CE 26, and PHA 4179 Toutatis (2 nights). The first 5 nights were observed using the old HPC1 camera (asteroids 35396, 40267 and 65803), while the last three nights with the new ST-9 camera (asteroids 2002 CE26 and 4179 Toutatis).

Unfortunately, two batches including the NEA 40267 were rejected by MPC, due to their large O-C residuals (up to about 5 arsec). Why did this happen? Following some afterward testing, we have realized that the time stamp was the problem. The HPC1 acquisition software "frozen" the PC clock for the period of exposure and reading process, thus the observing time was recorded delayed in the FITS headers. Because WinSNTP adjusts the correct time once every minute, the "freezing" effect delayed the PC clock by up to one minute. As we took exposures with a frequency less than one minute for most of the time these nights, the large proper motion of the asteroid (8-9 "/min) combined with the PC time "freezing" resulted in big residuals in the reduced positions. We were not aware before these two nights about this problem (happening only with the old HPC1 camera under Windows 95). Later, in order to correct this error for the HPC1, we have reset manually the PC time clock using WinSNTP after each image acquisition, in order to correctly reset the time before a new image. This unfortunate error is a warning, stressing once again the importance of the accurate time stamp necessary to observe fast moving sources.

*Table 2 -* The observing log of our NEA observations

| NEA/PHA (1) | Date (UT) (2) | Nr. pos. (3) | Inter. (min) (4) | Ref stars & catalogs (5) | ("/s) (6) | Mag (7) | Bands (8) | Exp (s) (9) | MPC status (10) |
|---|---|---|---|---|---|---|---|---|---|
| PHA (35396) | Nov 09 2002 | 24 | 50 | 2 (1 HIP, 1 2MASS) | 0.13 | R 13.1 V 13.8 | V, R, I | 20 | A |
| NEA (40267) | Feb 08 2003 | 9 | 54 | 2-5 (4 GSC 2.2, 1 Tycho-2) | 0.16 | R 13.1 V 13.3 | R, I | 20 | R |
| NEA (40267) | Feb 10 2003 | 16 | 229 | 2-6 (5 2MASS, 1 Tycho-2) | 0.13 | R 13.2 V 13.4 | V, R, I | 20 | R |
| PHA (65803) Didymos | Nov 21 2003 | 12 | 67 | 1-2 (Tycho-2) | 0.12 | R 12.4 V 12.7 | R, I | 30 | A |
| PHA (65803) Didymos | Dec 04 2003 | 35 | 82 | 3 (GSC 2.2) | 0.05 | R 12.9 V 13.2 | R, I | 30 | A |
| NEA 2002 CE26 | Aug 22 2004 | 15 | 84 | 3-6 (USNO-A2) | 0.06 | R 15.1 V 15.5 | R | 60 | U |
| PHA (4179) Toutatis | Aug 22 2004 | 21 | 131 | 2 (Tycho-2) | 0.01 | R 12.4 V 13.1 | R | 20 | A |
| PHA (4179) Toutatis | Sep 12 2004 | 20 | 55 | 2 (1 Tycho-2, 1 DENIS) | 0.02 | R 11.3 V 11.9 | B,V, R, I | 30 | A |

Table 2 gives the observing log, including the following columns: (1) – the MPC name of the object and its class (NEA or PHA), (2) – date in UT, (3) – number of positions observed, (4) – observed time interval in minutes, (5) – number of astrometric reference stars used and their catalog source, (6) – asteroid proper motion in arc seconds per second, (7) – the apparent magnitude measured in R and quoted by HORIZONS ephemerides in V, (8) – the observed bands, (9) – the exposure time in seconds, (10) – the MPC status of our batch (A = accepted, R = rejected, U = unknown).

Additional to observations in Table 2, two other nights were spent on two fainter asteroids, namely NEA 25330 (1999 KV 4) on Dec 2/3, 2002 (V=14.6) and NEA 52387 (1993 OM 7) on Jan 6/7, 2003 (V=14.9). Both of them could not be seen. Including these two attempted observations, the total number of nights spent on this project is ten.

## 6. DATA REDUCTION

### 6.1. IMAGE PROCESSING

The image processing consists in corrections for the bias, dark and flat field. We perform this in IRAF 2.12.2 under Linux. IRAF is distributed by the National Optical Astronomy Observatories, which are operated by the Association of Universities for Research in Astronomy, Inc., under cooperative agreement with the National Science Foundation (NOAO, 2004).

Dome flats were observed in the past, but various gradients were remarked, probably due to the small distance to the dome screen, also its small size. Given the excessive light pollution in Toronto, we obtained most of our flat fields by combining sets of 7-9 night sky images exposed 60-90s, with successive fields nodded arbitrarily in order to eliminate the stars, combining the frames using the median mode. Nevertheless, twilight flat fields are to be preferred.

With the former HPC1 camera, sets of about 11-15 bias frames and 7-9 dark frames exposed the same as the science frames were taken usually at the end of observing nights. With the new ST-9 camera, we subtracted from each science image and flat field frame the dark taken the moment of their acquisition.

### 6.2. ASTROMETRY

The major problem with the small field astrometry acquired from a light polluted place is the small number of reference stars in most images. Based on the number of reference stars present in the science field, at least two reduction methods are possible.

#### 6.2.1. Sky to Plate Transformation

Given two reference stars having the equatorial coordinates $(\alpha_1, \delta_1), (\alpha_2, \delta_2)$ and the CCD rectangular coordinates $(x_1, y_1), (x_2, y_2)$, then the coefficients $(a, b)$ expressing the rotation of the CCD with respect to the equator can be derived from the following equations:

$$a = \frac{x_2 + y_2}{(x_2 - x_1)^2 + (y_2 - y_1)^2}$$
$$b = \frac{x(y_2 - y_1) - y(x_2 - x_1)}{(x_2 - x_1)^2 + (y_2 - y_1)^2}$$

(3)

Here *x* and *y* are two constants, given by the classic formula expressing the sky-plate transformation (e.g., Montenbruck and Pfleger 1991):

$$x = \frac{\cos\delta_2 \sin(\alpha_2 - \alpha_1)}{\cos\delta_1 \cos\delta_2 \cos(\alpha_2 - \alpha_1) + \sin\delta_1 \sin\delta_2}$$

$$y = -\frac{\sin\delta_1 \cos\delta_2 \cos(\alpha_2 - \alpha_1) - \cos\delta_1 \sin\delta_2}{\cos\delta_1 \cos\delta_2 \cos(\alpha_2 - \alpha_1) + \sin\delta_1 \sin\delta_2}$$

(4)

With $(a, b)$ known, the asteroid equatorial coordinates $(\alpha_{ast}, \delta_{ast})$ can be found from the following equations:

$$\alpha_{ast} = \alpha_* + \arctan\left(\frac{x_{eq}}{\cos\delta_* - y_{eq} \sin\delta_*}\right)$$

$$\delta_{ast} = \delta_* + \arcsin\left(y_{eq} - x_{eq} \sin\delta_* \tan\left(\frac{\arctan(\alpha_{ast} - \alpha_*)}{2}\right)\right)$$

(5)

Here $(x_{eq}, y_{eq})$ represent the rectangular coordinates of the asteroid in the rotated system:

$$x_{eq} = a(x_{ast} - x_*) + b(y_{ast} - y_*)$$
$$y_{eq} = -b(x_{ast} - x_*) + a(y_{ast} - y_*)$$

(6)

### 6.2.2. Only One or Two Reference Stars in the Field

If only one or two reference stars are available in the field, at least one is recommended to be from Tycho-2 catalog. In this case, one can apply the sky to plate transformation as above.

In the rare case when only one Tycho-2 star is available in the field, a nearby "orientation field" which contains at least two Tycho-2 stars must be observed, from which we can get the plate solution, i.e., the parameters $(a, b)$. In order to define the same camera orientation, the orientation field must be chosen very close in space to the science frame. Then $(a, b)$ can be used with the science field and the catalog star as the reference, in order to derive the asteroid position.

If two reference stars (preferably of Tycho-2) are available in the science field, then the asteroid coordinates are determined by averaging its coordinates derived using both stars, individually. In this case, $(a, b)$ are determined from the same field.

Using this method and an ideal transformation equation (i.e. an absolute rectangular linear CCD perpendicular to the focal axis), the expected errors in the final positions are given by the statistics of the stellar positions, quoted for each star in its catalog entry.

We used our ORIENT and REL3 software (Vaduvescu 2004a) to calculate the field orientation and to reduce the equatorial positions, respectively.

### 6.2.3. Minimum Three Reference Stars in the Field

In this case, one can use IRAF (NOAO 2004) to perform the astrometry. First, CCMAP should be employed to compute the plate solution using the sky projection geometry using the default values. Second, CCTRAN is used to transform the asteroid coordinate list using the CCMAP plate solution.

Alternatively, one can use the same method from 6.2.2 to average the asteroid positions determined using all stars individually. In this case, the plate constants are derived from the same field, by averaging all available Tycho-2 pair results.

In both cases, Tycho-2 and 2MASS positions are to be preferred first, followed by USNO-A2, USNO-B1 and GSC 2.2. Given the scarcity of stars in our small field and the low magnitude limit, we prefer to use star positions taken from different catalogs, provided all are given in the same ICRS/J2000 system. Should enough stars are available, the individual positions with residuals greater than 0.3 arcsec than the average can be rejected.

### 6.3. PHOTOMETRY

No attempt was made to observe photometric standard stars on any of our observing nights, because of some pointing problems, risk of loosing the target, some non-photometric conditions, also because of the lack of observers (I have observed alone for most of the time).

Differential photometry using the catalog stars in the field was sufficient to provide first order magnitude estimates for the NEAs. The stellar magnitudes were cross-correlated and weighted based on the best available catalogs providing photometry, e.g., searched using VizieR (CDS 2004b). The individual differences derived from all available stars in the field were averaged, in order to get the apparent magnitude of the asteroid. We used the PHOT task of IRAF to measure the photometry.

## 7. COMPARISON WITH THE EPHEMERIDES

One can use the classic "observed minus calculated" (O-C) approach to check the astrometry quality before sending each data batch to MPC. To do so, we employed HORIZONS ICRF/J2000 topocentric ephemerides run close to the date we produced our data (usually a few days following the observations). In order to calculate the O-C residuals, we used our own INTER software which interpolates linearly our observations using the HORIZONS one minute step ephemerides.

An alternative to using INTER would be an online server which calculates residuals of asteroid positions using OrbFit ephemerides (Scvarc 2004). This service expect the client/user to upload a file including the observation batch entered in the usual MPC format, then calculates the O-Cs, their averages, returning also the plots with the scatter of residuals and residual distributions.

We present in Figure 2 the O-C values of our observations, run with INTER ephemerides interpolator based on the HORIZONS output. As one can see, most of the plots show residual values less than 0.3 arcsec, with no systematic trend with the UT. Only PHA 40267 (the batch rejected by MPC), shows the largest residuals and some systematic trend, due to the time stamp problem. The second observation of PHA (4179) Toutatis shows smaller residuals that the first, possibly due to its improved orbit, as more data became available closer to opposition (about 350 observations between our two observing dates, cf. with NEODyS). As one can see, the errors of the average O-C residuals are comparable with the combined positional errors of the stars employed for astrometry, listed in catalogs.

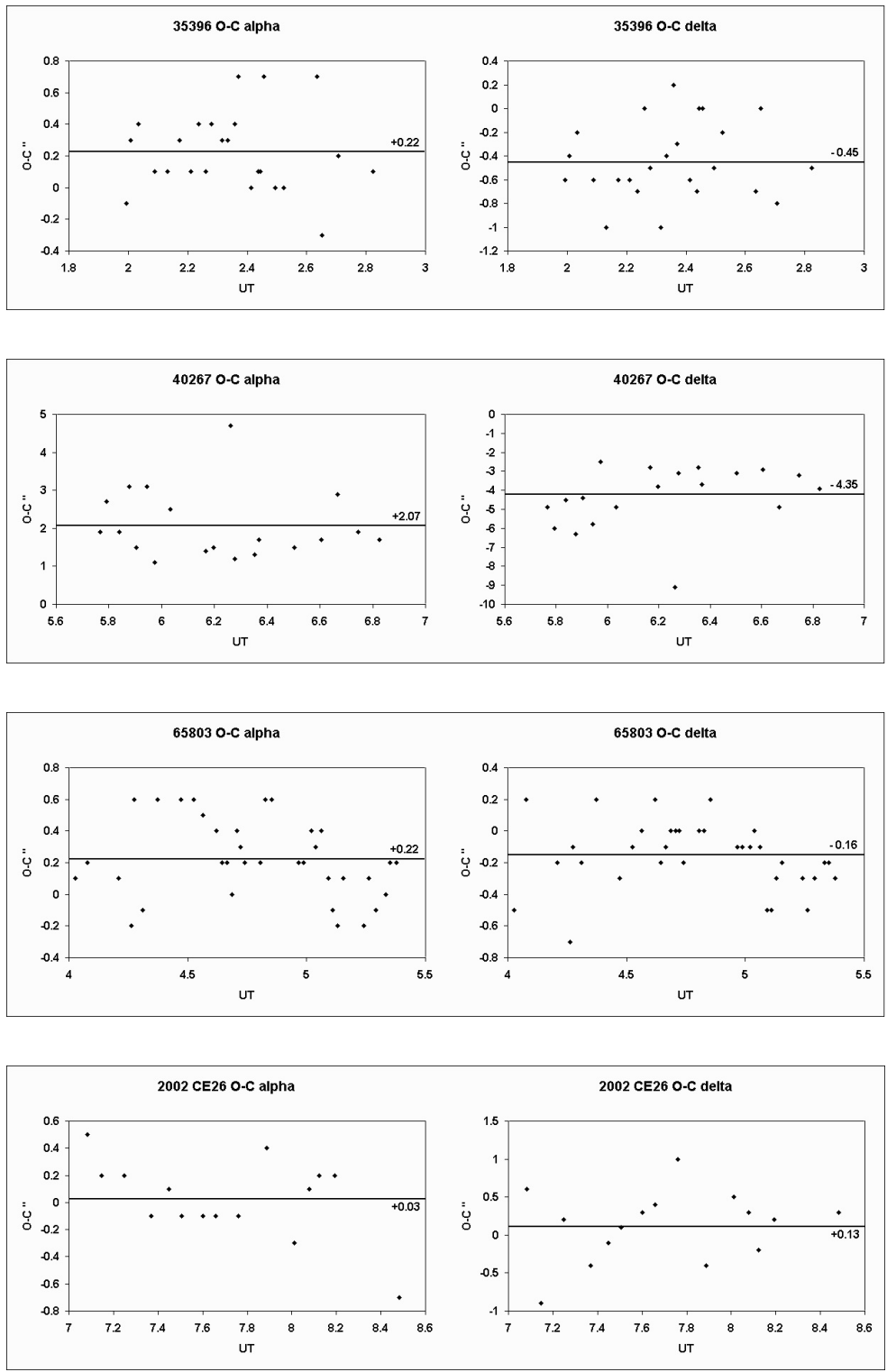

Fig. 2 – Please see caption on the next page.

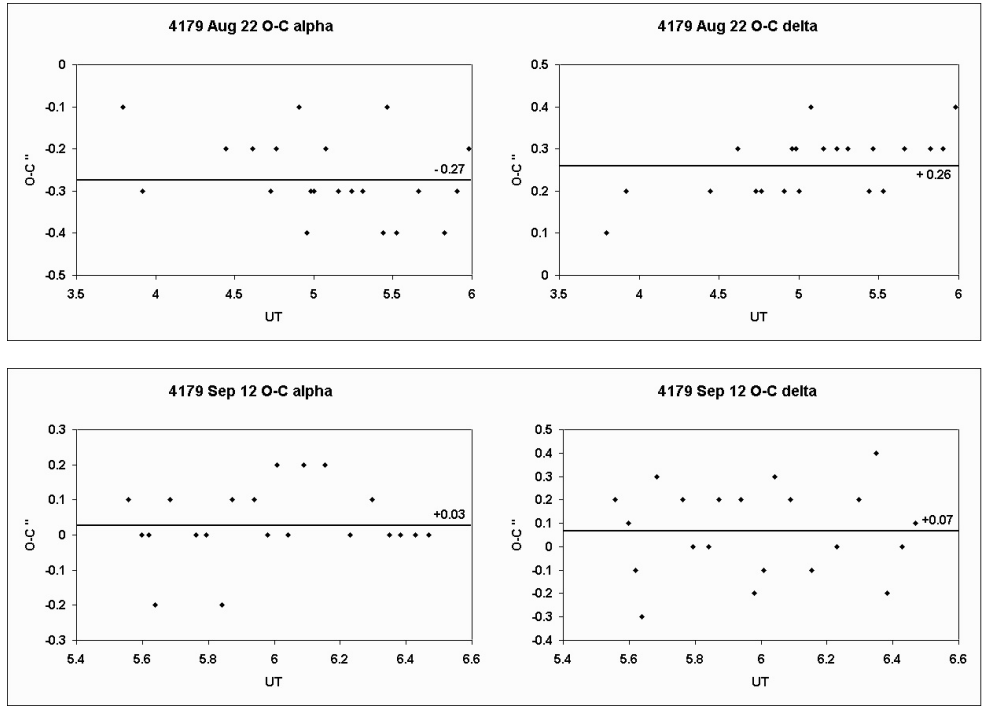

Fig. 2 – The Observed minus Calculated (O-C) values giving the residuals between our NEAs reduced data and the HORIZONS ephemerides, plotted in right ascension (left panels) and declination (right panels). The average values are plotted as a horizontal line and included to the right.

## 8. REPORTING THE DATA TO THE MPC

After reducing the data and checking the O-C residuals, we report by email the observations to Minor Planet Centre, using their strict specific format (MPC 2004a). As an example, we give bellow such a batch (including 4179 Toutatis observed on August 21/22, 2004).

```
COD H79
CON O. Vaduvescu, York University, Department of Physics and Astronomy, 128
CON Petrie Science Building, 4700 Keele Street, Toronto, ON, M3J 1P3, Canada
CON [ovidiuv@yorku.ca]
OBS O. Vaduvescu
MEA O. Vaduvescu
TEL 0.60-m f/13.6 Cassegrain + CCD SBIG ST-9
NET Tycho-2 (astrometry), USNO-B1 (photometry)
NUM 21

4179         C2004 08 22.15811 22 4 36.88 -12 41 12.2          12.5 R     H79
4179         C2004 08 22.16318 22 4 36.69 -12 41 13.7          12.4 R     H79
4179         C2004 08 22.18521 22 4 35.95 -12 41 20.8          12.4 R     H79
4179         C2004 08 22.19229 22 4 35.71 -12 41 23.0          12.4 R     H79
4179         C2004 08 22.19704 22 4 35.54 -12 41 24.6          12.4 R     H79
4179         C2004 08 22.19848 22 4 35.50 -12 41 25.1          12.4 R     H79
4179         C2004 08 22.20442 22 4 35.30 -12 41 27.0          12.4 R     H79
4179         C2004 08 22.20645 22 4 35.21 -12 41 27.6          12.4 R     H79
4179         C2004 08 22.20752 22 4 35.18 -12 41 27.9          12.4 R     H79
4179         C2004 08 22.20841 22 4 35.15 -12 41 28.2          12.4 R     H79
4179         C2004 08 22.21155 22 4 35.05 -12 41 29.0          12.4 R     H79
4179         C2004 08 22.21479 22 4 34.93 -12 41 30.2          12.4 R     H79
4179         C2004 08 22.21828 22 4 34.81 -12 41 31.3          12.4 R     H79
4179         C2004 08 22.22113 22 4 34.71 -12 41 32.2          12.4 R     H79
4179         C2004 08 22.22653 22 4 34.52 -12 41 33.9          12.5 R     H79
4179         C2004 08 22.22756 22 4 34.50 -12 41 34.2          12.5 R     H79
4179         C2004 08 22.23022 22 4 34.39 -12 41 35.1          12.4 R     H79
4179         C2004 08 22.23587 22 4 34.20 -12 41 36.8          12.4 R     H79
4179         C2004 08 22.24266 22 4 33.96 -12 41 38.9          12.4 R     H79
4179         C2004 08 22.24600 22 4 33.85 -12 41 39.9          12.5 R     H79
4179         C2004 08 22.24921 22 4 33.75 -12 41 40.8          12.4 R     H79
```

Should any problem exists with the reported data, the observer will receive a brief reply, otherwise probably the batch is accepted (the procedure is not clear). If accepted, usually within one day the batch will appear in the NEODyS online database (Milani et al 2004a), but the protocol again is not clear. For some time, MPC started to provide online archives with the front page and observation summary pages from each batch of their circulars, in order to reference the various observer contributions (MPC 2004b). We found two such circulars in this archive, listing some of our reported data including 80 positions of 3 NEAs (Vaduvescu 2004b).

Following the third asteroid batch we sent, York University Observatory has been assigned the observatory code H79 by the MPC. This probes both the quality of the data and the site location (as the reduced data sent to MPC are topocentric).

## 8. CONCLUSIONS

The number of NEAs and PHAs has increased dramatically in the last decade. This increase is due mainly to seven dedicated NEAs discovery programs, carried out mostly in the US, also due to some other efforts. All these dedicated programs work in the direction recommended by Spaceguard Survey and NASA to discover and characterize 90% of the existent NEO population with diameters larger than 1km by 2008.

After the discovery of a NEA, the next important step is the recovery at the next opposition (as the object gets too faint to be observed by any available facility). Then, the object must be followed up with additional observations in order to secure and improve its orbit, and finally to predict any collision threat in the future. As fainter NEAs must be recovered and tracked only by larger telescopes, the brighter ones can be observed using smaller facilities.

Even from a light polluted place, using a small telescope equipped with a CCD camera sub-sampling the seeing, the largest NEAs and PHAs with sizes in the range of ~1km or larger can be observed at opposition, close passing the Earth. An accurate time stamp must be available, which is especially important for the close encounters which produce very fast apparent motions. The maximum exposure time is given by both the asteroid apparent magnitude and its proper motion.

Although a large field of view and a large CCD format are recommendable in order to have at least five reference stars in the field (preferably of Tycho-2 catalog), even a smaller FOV can serve the astrometry, employing even less than three stars having accurate J2000 positions, such as those taken from denser catalogs as Tycho-2, 2MASS, USNO-B2 or GSC 2.2.

Between Nov 2002 and Sep 2004 we carried out a NEAs follow-up program using the 60-cm Cassegrain telescope at York University Observatory. We first used an HPC1 512x512 pixels camera (FOV 4.8 arcmin), which could reach a limiting magnitude V~14 in one minute, under the very light polluted sky of Toronto. This was replaced recently by a more sensitive SBIG ST-9 camera (FOV 3.5 or 5.5 arcmin using a focal lens reducer) which can reach V~16 in one minute. Between 2002 and 2004, we spent 10 nights on this project and got data for five NEAs at opposition, among which three objects were PHAs. The astrometry and differential photometry has been reduced using the available reference stars in the field, providing an estimated accuracy of about 0.1 arcsec and 0.1 mag, respectively. The observed minus calculated (O-C) residuals were less than 0.3 arcsec. The reduced data was sent to Minor Planet Centre (MPC). Four batches were accepted and included in NEODyS database and MPC circulars, and a code for York Observatory was assigned, H79.

*Acknowledgments.* Prof. Paul Delaney, the Director of York University Observatory, has supported me with this project. Also, Paul and the undergraduate student Brenda Shaw assisted in two nights with the observations. During 2004, I have discussed this project and exchange some information with Alin Nedelcu of the Astronomical Institute of the Romanian Academy, who is involved in a similar NEA follow-up project using the double 38cm diameter refractor in Bucharest. Dr. Jerome Berthier and Dr. Mirel Birlan of IMCCE in Paris, France helped me to compare my first data reduction batch versus MIDAS.